\documentstyle[aps,epsfig,prb]{revtex}
\begin{document}

\title{Orbital currents and
cheap vortices in underdoped cuprates}

\author{Patrick A. Lee}

\address{Center for Materials Science and Engineering and
Department of Physics, MIT, Cambridge, MA 02139 USA}

\date{\today}

\maketitle

\begin{abstract}
In the past several years, we have developed a theory for the underdoped
cuprates in collaboration with X.-G. Wen, based on an SU(2) formulation
of the $t$-$J$ model.  In this formulation, the staggered flux state plays
a central role as the progenitor of the N\'{e}el state at half-filling
and a close competitor to the $d$-wave superconductor with small doping
where it is characterized by staggered orbital currents.
We found support for this point of view when we discovered
fluctuating orbital currents in the Gutzwiller projected BCS
wavefunction.  We shall argue that low-energy vortices, where the
staggered flux state is stabilized in the core, are needed to explain
many of the unusual properties of underdoped cuprates.  Proposed
experiments to look for these orbital currents will be discussed.
\end{abstract}

The phenomenon of high temperature superconductivity in cuprates is 
generally agreed to be associated
with doping into a Mott insulator.  The undoped material is an 
antiferromagnet with a large
exchange energy $J$ of order 1500~K.  The doped holes hop with a 
matrix element $t$, which is
estimated to be approximately $3J$.  However, the  N\'{e}el state is 
not favorable for hole
hopping, because after one hop the spin finds itself in a 
ferromagnetic environment.  Thus it
is clear that the physics is that of competition between the exchange 
energy $J$ and the hole
kinetic energy per hole $xt$.  Apparently the superconducting state 
emerges as the best
compromise, but how and why this occurs is the central question of 
the high T$_c$ puzzle.  In
the underdoped region this competition results in physical properties 
that are most anomalous.
The metallic state above the superconducting T$_c$ behaves in a way 
unlike anything we have
encountered before.   Essentially, an energy gap appears in some 
properties and not others, and
this metallic state has been referred to as the pseudogap state.  We 
will focus our attention
on this region because the phenomenology is well established and we 
have the best chance of
sorting out the fundamental physics.  Indeed, I will argue that a 
consensus is emerging on the
framework needed to describe this phenomenon and that detailed 
proposals are amenable to
experimental tests.

The pseudogap phenomenon is most clearly seen in the uniform 
susceptibility.  For example,
Knight shift measurement in the YBCO 124 compound shows that while 
the spin susceptibility is
almost temperature independent between 700~K and 300~K, as in an 
ordinary metal, it decreases
below 300~K and by the time the T$_c$ of 80~K is reached, the system 
has lost 80\% of the spin
susceptibility.\cite{1}  Similarly, the linear $T$ coefficient of the 
specific heat shows a marked
decrease below room temperature.  It is apparent that the spins are 
forming into singlets and
the spin entropy is gradually lost.  On the other hand, the frequency 
dependent conductivity
behaves very differently depending on whether the electric field is 
in the $ab$ plane
$(\sigma_{ab})$ or perpendicular to it $(\sigma_c)$.  At low 
frequencies (below 500~cm$^{-1}$)
$(\sigma_{ab})$ shows a typical Drude-like behavior for a metal with 
a width which decreases
with temperature, but an area (spectral weight) which is independent 
of temperature.\cite{2}  Thus
there is no sign of the pseudogap in the spectral weight.  This is 
surprising because in other
examples where an energy gap appears in a metal, such as the onset of 
charge or spin density
waves, there is a redistribution of the spectral weight from the 
Drude part to higher
frequencies.  On the other hand below 300~K $\sigma_c(\omega)$ is 
gradually reduced below
500~cm$^{-1}$ and a deep hole is carved out of $\sigma_c(\omega)$ by 
the time T$_c$ is
reached.  Finally, angle-resolved photoemission shows that an energy gap (in
the form of a pulling back of the leading edge of the electronic 
spectrum from the Fermi
energy) is observed near momentum $(0,\pi)$ and the onset of 
superconductivity is marked by the
appearance of a small coherent peak at this gap edge.

The pseudogap phenomenology is well explained by a cartoon picture 
which emerges from the RVB
(resonating valence band) theory of Anderson.\cite{3}  The spins are 
paired into singlet pairs.
However, the pairs are  not static but are fluctuating due to quantum 
mechanical superposition,
hence the term quantum spin liquid.  The singlet formation explains 
the appearance of the spin gap and the
reduction of spin entropy.  The doped holes appear as vacancies in 
the background of singlet
pair liquid and can carry a current without any energy gap.  However 
in $c$-axis conductivity
and electron is removed from one plane and placed on the next.  The 
intermediate state is an
electron which carries spin 1/2 and therefore it is necessary to 
break a singlet pair and pay
the spin-gap energy.  The same consideration applies to the 
photoemission experiment.  Finally, according to RVB
theory, superconductivity emerges when the holes become phase 
coherent.  The spin singlet
familiar in the BCS theory has already been formed.

While the above picture is appealing, there has been another popular 
explanation which at first
sight sounds less exotic.  The idea is that the pseudogap 
phenomenology can be understood as a
superconductor with robust amplitude but strong phase fluctuation. 
The superfluid density
$\rho_s$ which controls the phase stiffness is proportional to the 
doping concentration $x$ and
becomes small in the underdoped region.  As emphasized by 
Uemura\cite{4} and by Emery and Kivelson,\cite{5}
T$_c$ is controlled by $\rho_s$ and is much lower than the energy 
gap.  We shall now argue that
these two points of views are not unrelated to each other.  Setting 
aside the question of where
the strong pairing amplitude comes from in the first place, that the 
phase fluctuation scenairo
is incomplete can be seen from the following argument.  In two 
dimensions the destruction of
superconducting order is via the Berezinskii-Kosterlitz-Thouless 
(BKT) theory of vortex
unbinding.  Above T$_c$ the number of vortices proliferate and the 
normal metallic state is
reached only when the vortex density is so high that the cores 
overlap.  (There is considerable
lattitude in specifying the core radius, but this does not affect the 
conclusion.)  At lower
vortex density, transport properties will resemble a superconductor 
in the flux flow regime.
In ordinary superconductors, the BKT temperature is close to the mean 
field temperature, and
the core energy rapidly becomes small.  However, in the present case, 
it is postulated that the
mean field temperature is high, so that a large core energy is 
expected.  Indeed, in a
conventional core the order parameter and energy gap vanish, costing 
$\Delta^2_0/E_F$ per unit area of energy.
Using a core radius of $\xi = V_F/\Delta_0$, the core energy of a 
conventional superconductor is $E_F$.  In our
case, we may replace $E_F$ by $J$.  If this were the case, the 
proliferation of vortices would not happen until a
high temperature $\sim J$ independent of $x$ is reached.  Thus for 
the phase fluctuation scenario to to work, it
is essential to have ``cheap'' vortices, with energy cost of order 
T$_c$.  Then the essential problem is to
understand what the vortex core is made of.  Put another way, there 
has to be a competing state with energy very
close to the $d$-wave superconductors which constitute the core.  The 
vortex core indeed offers a glimpse of the
normal state reached when $H$ exceeds
$H_{c2}$, and is an important constituent of the pseudogap state above T$_c$.

To summarize, the pseudogap phenomenology is so clearly defined by 
experiments that I believe a
broad consensus on the theoretical framework is emerging.  The 
following points are common to a
number of investigators.

\begin{enumerate}
\item The key issue is how holes are accommodated in a doped Mott insulator.

\item The superconducting state is characterized by a large energy 
gap and a small $\rho_s$.
The transition temperature is controlled by phase coherence.

\item The superconducting state does not evolve out of a normal state 
with well defined
quasiparticles.  Therefore the question of pairing mechanisms in the 
traditional sense of
identifying the boson being exchanged between quasiparticles is not 
the right question.

\item What is needed is to identify the competing state which lives 
iside the vortex core.

\end{enumerate}

What are the candidates for the competing order?  A candidate which 
has attracted a lot of
attention is the stripe phase.\cite{6,7}    In the LSCO familty, 
dynamical stripes (spin
density waves) are clearly important, especially near $x = {1\over 
8}$.  There are recent
report of incommensurate SDW nucleating around vortices.  However, 
until now there has been little
evidence for stripes outside of the LSCO family.  On the theoretical 
side, as a competing state
it is not clear how the stripes are connected to $d$-wave 
superconductivity and it is hard to
understand how the nodal quasiparticles turn out to be most sharply 
defined on the
Fermi surface, since these have to transverse the stripes at a 
$45^\circ$ angle.

A second candidate for the vortex core is the antiferromagnetic 
state.  This possibility was
first proposed several years ago in the context of the SO(5) 
theory.\cite{8}  This theory is
phenomenological in that it involves only bosonic degrees of freedom 
(the SDW and pairing order
parameters).  The quasiparticles are out of the picture and indeed 
this theory applies equally
well to $d$-wave or a fully gapped $s$-wave superconductor.  Thus the 
fundamental question of
how the holes are accommodated has not really been addressed.  The 
nucleation of SDW around the core in the
LSCO family lends support to the existence of antiferrmagnetic order 
inside the core.
Furthermore, there are reports of enhanced antiferromagnetic spin 
fluctuations, and perhaps even
static order, using NMR.\cite{9,10}  I shall argue next that other 
considerations also lead to
antiferromagnetic fluctuations and possibly static orders inside the 
vortex core, so that the
observation of antiferromagnetic cores does not necessarily imply the 
existence of SO(5)
symmetry.

Finally, I come to the candidate which we favor --- the staggered 
flux state with orbital
currents.\cite{11}  Indeed, Lee and Wen have successfully
constructed a ``cheap'' vortex state.\cite{12}  The staggered flux phase has an
advantage over other possibilities in that its excitation spectrum is 
similar to the $d$-wave
superconductor and the SU(2) theory allows us to smoothly connect  it to the
superconductivity.   We also regard the staggered flux phase as the 
precursor to  N\'{e}el
order, so that antiferromagnetic  fluctuations or even SDW order are 
accommodated naturally.
Of course, it is experiments which have the final say as to which 
candidate turns out to be
realized.  Our strategy is to work out as many experimental 
consequences as we can and propose
experiments to confirm or falsify our theory.

The staggered flux state was first introduced as a mean field 
solution at half-filling\cite{13}
and later was extended to include finite doping.\cite{14}   At 
half-filliing, due to the constraint of no double
occupation, the staggered flux state corresponds to an insulating 
state with power law decay in the spin
correlation function.  It is known that upon including gauge 
fluctuations which enforce the
constraint, the phenomenon of confinement and chiral symmetry 
breaking occurs, which directly
corresponds to  N\'{e}el ordering.\cite{15}  The idea is that with 
doping, confinement is
suppressed at some intermediate energy scale, and the state can be 
understood as fluctuating
between the staggered flux state and the $d$-wave superconducting 
state.  Finally, when the
holes become phase coherent, the $d$-wave superconducting state is 
the stable ground state.
Thus the staggered flux state may be regarded as the ``mother state'' 
which is an unstable
fixed point due to gauge fluctuations.  It flows to  N\'{e}el 
ordering at half-filling and to
the $d$-wave superconductor for sufficiently large $x$.  Thus the 
staggered flux state plays a
central role in this kind of theory.  We should point out that the 
staggered flux state (called
the $D$-density wave state) has recently been proposed as the ordered 
state in the pseudogap
region.\cite{16}  As explained elsewhere,\cite{17} we think that this 
view is not supported
by experiment and we continue to favor the fluctuation picture.

The above picture finds support from studies of  projected 
wavefunctions, where the
no-double-occupation constraint is enforced by hand on a computer. 
With doping the best state
is a projected
$d$-wave state.   Recently we calculated the current-current 
correlation function of this state
$
c_j(k,\ell) = < j(k)j(\ell) >
$
where $j(k)$ is the physical electron current on the bond $k$.  The 
average current $<j(k)>$ is
obviously zero, but the correlator exhibits a staggered circulating 
pattern.\cite{18}
Such a pattern is absent in the $d$-wave BCS state before
projection, and is a result of the Gutzwiller projection.  Our result 
for $c_j$ is consistent with exact
diagonalization of two holes in 32 sites.\cite{19}

The staggered current generates a staggered physical
magnetic field (estimated to be 10--40 gauss)\cite{14,18} which may 
be detected, in principle, by neutron
scattering.  In practice the small signal makes this a difficult, 
though not impossible experiment and we
are motivated to look for situations where the orbital current may 
become static or quasi-static.  Recently, we
analyzed the structure of the $hc/2e$ vortex in the superconducting 
state within the SU(2) theory and concluded
that in the vicinity of the vortex core, the orbital current becomes 
quasi-static, with a time scale determined by
the tunnelling between two degenerate staggered flux states.\cite{12} 
It is very likely that this time is
long on the neutron time scale.   Thus we propose that a quasi-static 
peak centered around $(\pi,\pi)$ will appear
in neutron scattering in a magnetic field, with intensity 
proportional to the number of vortices.  The time
scale may actually be long enough for the small magnetic fields 
generated by the orbital currents to be
detectable by
$\mu$-SR or Yttrium NMR.  Again, the signal should be proportional to 
the external fields.  (The NMR experiment
must be carried out in 2--4--7 or 3 layer samples to avoid the 
cancellation between bi-layers.)  We have also
computed the tunnelling density of states in the vicinity of the 
vortex core, and predicted a rather specific kind
of period doubling which should be detectable by atomic resolution 
STM.\cite{20}  The recent
report\cite{21} of a static field of $\pm 18$ gauss in underdoped 
YBCO which appears in the vortex state is
promising, even though muon cannot distinguish between orbital 
current or spin as the origin of the magnetic
field.  We remark that in the underdoped antiferromagnet, the local 
moment gives rise to a field of 340 gauss
at the muon site.  Thus if the 18 gauss signal is due to spin, it 
will correspond to roughly $1/20$th of the
full moment.

We remark that our analytic model of the vortex core is in full 
agreement with the numerical solution of
unrestricted mean field $\Delta_{ij}$ and $\chi_{ij}$ by Wang, Han 
and Lee.\cite{22} Recently we found that for
small doping in the $t$-$J$ model a small moment SDW co-exists with 
orbital currents in the vortex core.\cite{23}
More generally, we expect $(\pi,\pi)$ spin fluctuations to be 
enhanced \cite{24} so that the tendency to
antiferromagnetism is fully compatible with the staggered flux 
picture.  This vortex solution is also interesting
in that the tunnelling density of states show a gap, with no sign of 
the large resonance associated with
Caroli-deGennes-type core levels found in the standard BCS model of 
the vortex. There exists a single bound state
at low lying energy,\cite{23} in agreement with STM experiments.
  The low density of states inside the vortex core has an important 
implication.  In the standard
Bardeen-Stephen model of flux-flow resistivity, the friction 
coefficient of a moving vortex is due to dissipation
associated with the vortex core states.  Now that the core states are 
absent, we can expect anomalously small
friction coefficients for underdoped cuprates.  The vortex moves fast 
transverse to the current and gives rise to
large flux-flow resistivity.   Since the total conductivity is the 
sum of the flux-flow conductivity and the
quasiparticle conductivity, it is possible to get into a situation 
where the quasiparticle conductivity dominates
even for H
$\ll$ H$_{c2}$.   Thus the ``cheap'' and ``fast'' vortex opens the 
possibility of having vortex states above the
nominal T$_c$ and H$_{c2}$, when the resistivity looks like that of a 
metal, with little sign of flux-flow
contribution.  From this point of view, the large Nerst effect 
observed by Ong and
co-workers \cite{25} over a large region in the H-T plane above the 
nominal T$_c$ and H$_{c2}$ (as determined
by resistivity) may be qualitatively explained.

Finally, I would like to mention my recent work with N. Nagaosa on 
new collective modes in the superconductor
state. \cite{26}  In ordinary superconductors, there is a single 
pairing order parameter $\Delta$ and associated
with it are the amplitude mode and the phase modes.  In our theory 
the hopping matrix element $\chi_{ij}$ is also a
dynamical degree of freedom.  The basic reason is that the hole 
hopping amplitude depends sensitively on the
spin configuration, as we explained earlier, so that $\chi_{ij}$ 
encodes the effect due to the spin degree
of freedom.  With the pairing and hopping amplitudes as dynamic 
degrees of freedom, it is natural that we
can expect new collective modes.  These can be evaluated in a direct 
way within mean field theory in analogy
with the standard RPA treatment, but the SU(2) theory allows us to 
classify them.  We find two new classes of
collective modes.  The first is called the $\theta$-mode, which 
describes the local fluctuation towards the
staggered flux phase, and has the effect of generating the orbital 
current correlation described earlier.
The second is called the $\phi$-gauge mode.  This mode involves the 
relative oscillation of the
{\em amplitudes} of $\chi_{ij}$ and $\Delta_{ij}$ in such a way that 
$|\chi_{ij}|^2 + |\Delta_{ij}|^2$ is
constant, and this oscillation is modulation in a staggered way. 
Thus this mode is most important near
$(\pi,\pi)$, but it  disperses throughout $k$ space in a way which we 
have computed.  The frequency of the
mode is estimated to be of order $J$ for $x = 0.15$.  This mode 
couples to Raman scattering in the standard
   way so that optical Raman scattering may observe the mode at 
$\mbox{\bf{k}} = 0$.  X-Ray Raman scattering should
be able to map out the entire dispersion, but higher energy 
resolution than what is currently available will
be needed.  It turns out the LSCO offers a special opportunity to 
couple to this mode.  In the LTO phase the
copper oxygen bonds are buckled in a staggered way.  A uniform motion 
of the oxygen along the $c$-axis leads
to a staggered modulation of the hopping matrix element.  Thus we 
predict that a new collective mode at
$(\pi,\pi)$ will show up as a side band of the buckling mode in 
$c$-axis conductivity.  Unfortunately, the
relative spectral weight of one side band is estimated to be very 
small (of order $5 \times 10^{-3}$),
making its detection a challenging one for experimentalists.

I thank X.-G. Wen, N.  Nagaosa, D. Ivanov, J. Kishine and Y. Morita 
for their collaboration on the
work reviewed here.  I acknowledge the support of NSF through the 
MRSEC program grant no.
DMR98--08941.

\end{document}